%% file: paper.tex
\documentclass[11pt, letterpaper]{article}
\usepackage{amsmath,amssymb,amsthm}
\usepackage{booktabs}
\usepackage{diagbox}
\usepackage{enumitem}
\usepackage[margin=1in]{geometry}
\usepackage{graphicx}

\input{def}

\title{Subset Sampling and Its Extensions} 

\author{
	Jinchao Huang, Sibo Wang \\[3mm]
	The Chinese University of Hong Kong \\
	Hong Kong, China \\
	{\em \{jchuang, swang\}@se.cuhk.edu.hk}
}

\begin{document}

\maketitle

\begin{abstract}
    This paper studies the \emph{subset sampling} problem. The input is a set $\mathcal{S}$ of $n$ records together with a function $\textbf{p}$ that assigns each record $v\in\mathcal{S}$ a probability $\textbf{p}(v)$. A query returns a random subset $X$ of $\mathcal{S}$, where each record $v\in\mathcal{S}$ is sampled into $X$ independently with probability $\textbf{p}(v)$. The returned subset must be independent of those returned by the previous queries. The goal is to store $\mathcal{S}$ in a data structure to answer queries efficiently.

    If $\mathcal{S}$ fits in memory, the problem is interesting when $\mathcal{S}$ is dynamic. We develop a dynamic data structure with $\mathcal{O}(1+\mu_{\mathcal{S}})$ expected \emph{query} time, $\mathcal{O}(n)$ space and $\mathcal{O}(1)$ amortized expected \emph{update}, \emph{insert} and \emph{delete} time, where $\mu_{\mathcal{S}}=\sum_{v\in\mathcal{S}}\textbf{p}(v)$. The query time and space of this data structure are optimal.

    If $\mathcal{S}$ does not fit in memory, the problem is difficult even if $\mathcal{S}$ is static. Under this scenario, we present an I/O-efficient algorithm that  answers a \emph{query} in $\mathcal{O}\left((\log^*_B n)/B+(\mu_\mathcal{S}/B)\log_{M/B} (n/B)\right)$ amortized expected I/Os using $\mathcal{O}(n/B)$ space, where $M$ is the memory size, $B$ is the block size and $\log^*_B n$ is the number of iterative $\log_2(.)$ operations we need to perform on $n$ before going below $B$.

    In addition, when each record is associated with a real-valued key, we extend the \emph{subset sampling} problem to the \emph{range subset sampling} problem, in which we require that the keys of the sampled records fall within a specified input range $[a,b]$. Again the returned subset must be independent of those returned by the previous queries. For this extension, we also provide a nontrivial solution under the dynamic setting, with $\mathcal{O}(\log n+\mu_{\mathcal{S}\cap[a,b]})$ expected \emph{query} time, $\mathcal{O}(n)$ space and $\mathcal{O}(\log n)$ amortized expected \emph{update}, \emph{insert} and \emph{delete} time.
\end{abstract}

\vspace{30mm}
\noindent {\bf Keywords:} Subset Sampling, External Memory, Dynamic Maintenance, Range Subset Sampling

\newpage

\input{sections/introduction.tex}

\input{sections/dynamic_subset_sampling}

\input{sections/em}

\input{sections/dynamic_range_subset_sampling}

\input{sections/conclusion}

\bibliographystyle{plainurl}
\bibliography{reference}

\end{document}

%% file: def.tex
\def\header{\vspace{2mm} \noindent}

\theoremstyle{plain}
\newtheorem{theorem}{Theorem}[section]

\newtheorem{lemma}[theorem]{Lemma}

\theoremstyle{definition}
\newtheorem{definition}[theorem]{Definition}

\theoremstyle{remark}

\newtheorem{problem}[theorem]{Problem}
\newtheorem{example}[theorem]{Example}

\newcommand\Bigo[1]{\mathcal{O}\left(#1\right)}
\newcommand\bigo[1]{\mathcal{O}(#1)}

\newcommand{\bin}[2]{\mathit{Binomial}(#1,#2)}
\newcommand{\geo}[1]{\mathit{Geometric}(#1)}
\def\integer{\mathbb{Z}}
\def\lc{\lceil}
\def\lf{\lfloor}
\renewcommand{\max}[2]{\mathrm{max}\{#1,#2\}}
\renewcommand{\min}[2]{\mathrm{min}\{#1,#2\}}
\def\rc{\rceil}
\def\rf{\rfloor}
\def\real{\mathbb{R}}
\newcommand\norm[1]{|#1|}
\newcommand\restr[2]{{
  \left.\kern-\nulldelimiterspace 
  #1
  \vphantom{|}
  \right|_{#2}
  }}
\newcommand{\set}[1]{\{#1\}}
\newcommand{\setht}[2]{\{#1,\ldots, #2\}}
\newcommand\ti{\times}
\newcommand{\unif}[2]{\mathit{Uniform}(#1,#2)}

\def\C{\mathcal{C}}
\def\D{\mathcal{D}}

\def\I{\mathcal{I}}
\def\P{\mathcal{P}}

\renewcommand{\S}{\mathcal{S}}
\def\T{\mathcal{T}}

\def\bp{\textbf{p}}
\def\bq{\textbf{q}}

\newcommand\sbp[0]{\widetilde{\textbf{p}}}

\newcommand\sbpo[1]{\widetilde{\textbf{p}}(#1)}
\newcommand\pv[0]{\textbf{p}(v)}
\newcommand\phv[0]{\hat{\textbf{p}}(v)}
\newcommand\ppv[0]{\textbf{p}^{\prime}(v)}
\newcommand\pppv[0]{\textbf{p}^{\prime\prime}(v)}
\newcommand\sX[0]{\widetilde{X}}
\newcommand\smS[0]{\widetilde{\mathcal{S}}}
\newcommand\mSab[2]{\mathcal{S}_{[#1,#2]}}

\newcommand{\que}[0]{\emph{query}}

\newcommand\upd[0]{\emph{update}}
\newcommand\updtwo[2]{\emph{update}(#1,#2)}
\newcommand\ins[0]{\emph{insert}}
\newcommand\instwo[2]{\emph{insert}(#1,#2)}
\newcommand\del[0]{\emph{delete}}
\newcommand\delone[1]{\emph{delete}(#1)}
\newcommand\compthree[3]{#1\leq#2\leq#3}

\newcommand\bigone[0]{\bigo{1}}
\newcommand\bigolog[0]{\bigo{\log n}}
\newcommand\bigonlog[0]{\bigo{n\log n}}
\newcommand\bigon[0]{\bigo{n}}

\newcommand\bigomu[0]{\bigo{1+\mu_\S}}

\newcommand\mul[0]{\mu_{S_l}}
\newcommand\twologone[0]{L-1}
\newcommand\twologtwo[0]{L}
\newcommand\pa{\frac{3}{16}} 
\newcommand\bpa{\frac{13}{16}} 
\newcommand\pb{\frac{7}{16}}
\newcommand\bpb{\frac{9}{16}}
\newcommand\pc{\frac{4}{16}}
\newcommand\bpc{\frac{12}{16}}
\newcommand\pd{\frac{5}{16}}
\newcommand\bpd{\frac{11}{16}}
\newcommand\qa[0]{\textbf{q}_{\bp^\prime}(\hat{\S})}
\newcommand\qb[0]{\textbf{q}_{\bp^{\prime\prime}}(\hat{\S})}

%% file: sections/introduction.tex
\section{Introduction}
    A typical query in Database Management Systems (DBMS) returns all records satisfying the input predicate. It works fine when the data size is small. Recent years, however, have seen an explosion in the size of data, which increases the query output size dramatically. For example, the output size of a query with selectivity $1\%$ on a dataset of size a trillion (this is only a moderate size in the era of big data) would be $10^{10}$. Reporting all these records incurs prohibitive time costs, especially when disk or network I/Os are involved.

    This motivates the study of \emph{query sampling}, a classic approach that was introduced to the database community in the 1990s. The goal of \emph{query sampling} is to return a random subset of the set of records satisfying the input predicate instead of the whole set, thus greatly reducing query response time. The importance of such a sampled subset has long been recognized, even in the dawn of the era of big data. Olken and Roten's survey~\cite{survey} published in 1995 excellently presents the benefits of \emph{query sampling}. The unprecedented gigantic data volume nowadays will only strengthen the importance of \emph{query sampling}. What's more, in many real-world applications, enquiring the entire query result is not compulsory, and a sampled subset already serves the analytic purpose well.

    This paper studies the \emph{subset sampling} (SS) problem, a type of \emph{query sampling} that was widely studied, e.g.,~\cite{constant_p_ss, logn+u_ss, 1+u_ss}. In \emph{subset sampling} query, each record is associated with a specified probability, and its chance of being sampled solely depends on this probability, independent of other records. When each record is associated with a real-valued key, the \emph{subset sampling} problem can be extended to the \emph{range subset sampling} (RSS) problem, in which we only sample from the set of records whose keys fall into the range $[a,b]$ given at query time. For instance, we may want to draw query samples from the records of products whose prices fall into a range known at query time. 

    In the literature, existing studies on \emph{subset sampling} ~\cite{constant_p_ss, logn+u_ss, 1+u_ss} focus on static datasets that reside in memory. We improve over existing studies by first designing a dynamic data structure for the setting where the datasets may change over time. Next, we further investigate how to handle \emph{subset sampling} under the external memory setting and design an I/O-efficient data structure for this problem. 
    In addition, we also design a dynamic data structure for the \emph{range subset sampling} problem we proposed above. We give formal problem definitions in Section \ref{sec:problem-definition}, review related works in Section \ref{sec:related-work}, and summarize our contributions in Section \ref{sec:contribution}.

    \input{sections/introduction/problem_definitions}

    \input{sections/introduction/related_work}

    \input{sections/introduction/contributions}

%% file: sections/introduction/problem_definitions.tex
\subsection{Problem Definitions}\label{sec:problem-definition} 
    Given a set $\S$ of records $\S=\setht{v_1,v_2}{v_n}$ together with a function $\bp$ that assigns each record $v\in\S$ a probability $\pv$, the \emph{subset sampling} query asks to randomly sample a subset $X$ of $\S$, where each $v\in\S$ is sampled into $X$ independently with probability $\pv$. The sampled subset must be independent of those returned by the previous queries. More formally, we have the following definition.
    \begin{problem}[Subset Sampling (SS)]
        Given a set $\S$ of $n$ records, and a function $\bp:\S\to[0,1]$, draw a random subset $X$ from $\S$ with
        $$
            \Pr[X=T\subseteq \S]=\left( \prod_{v\in T}\pv \right)\left( \prod_{v\in {\S\setminus T}}(1-\pv) \right).
        $$
    \end{problem}
    We call the tuple $\Psi=\left<\S,\bp\right>$ a \emph{subset sampling} problem instance. We denote the expected size of $X$, i.e., $E[|X|]=\sum_{v\in\S}\pv$, as $\mu_\S$.

    When each input record is associated with a real-valued key $k$ that we care about, the \emph{subset sampling} query can be extended to the \emph{range subset sampling} query as we may want to sample from the set of records whose key $k$ fall into a specific range instead of on the whole domain. In particular, given a range $[a,b]$ at query time, we only want to sample a subset from the set $\mSab{a}{b}=\{v|v\in\S, \compthree{a}{v.k}{b}\}$. For instance, a sales company might be interested in customers from a certain age group; the customers of Amazon might be interested in products from a certain price range. 
    Again the sampled subset must be independent of those returned by the previous queries. The formal definition is as follows.
    \begin{problem}[Range Subset Sampling (RSS)]
        Given a set $\S$ of $n$ records, where each record $v\in\S$ is associated with a real-valued key $v.k$, a function $\bp:\S\to[0,1]$, and a query range $[a, b]$, draw a random subset $Y$ from $\mSab{a}{b}=\{v|v\in\S, \compthree{a}{v.k}{b}\}$ with
        $$
            \Pr[Y=T\subseteq \mSab{a}{b}]=\left( \prod_{v\in T}\pv \right)\left( \prod_{v\in {\mSab{a}{b}\setminus T}}\left(1-\pv\right) \right).
        $$
    \end{problem}
    We call tuple $\psi=\left<\S,\bp\right>$ a \emph{range subset sampling} problem instance. We denote the expected size of $Y$, i.e., $E[|Y|]=\sum_{v\in\mSab{a}{b}}\pv$, as $\mu_{\S\cap[a,b]}$. 
    For ease of exposition, we abuse the notation of $v$ to represent $v.k$ when we discuss the \emph{range subset sampling} problem.

    We further define three dynamic operations: {\em (i)} $\updtwo{v}{p}$ that updates the probability $\bp(v)$ to $p$; {\em (ii)} $\instwo{v}{p}$ that inserts $v$ into $\S$ and sets $\pv$ to $p$; {\em (iii)} $\delone{v}$ that deletes $v$ from $\S$. Without loss of generality, we assume no duplicate keys exist in $\S$. These operations are collectively referred to as \emph{modification} operations. 
        
    \header{\bf Computation model.} 
        We study SS in both scenarios where the input set $\S$ fits or does not fit in memory, respectively. We study RSS in the scenario where the input set $\S$ fits in memory. 
        
        When the input set fits in memory, we discuss our algorithms on the \emph{Real RAM} (RR) model of computation~\cite{realram1,realram2}. In particular, we will assume that the following operations take constant time:
        \begin{itemize}
            \item Access a memory location.
            \item Generate a random value from the standard uniform distribution $\unif{0}{1}$ or generate a random integer in [0, $2^w-1$].
            \item Basic arithmetical operations involving real numbers like addition, multiplication, division, comparison, truncation, and evaluating any fundamental function like exp and log.
        \end{itemize}

        \noindent Under the \emph{Real RAM} model of computation, we can generate a random value $g$ from the geometric distribution $\geo{p}$ in $\bigone$ time~\cite{1+u_ss} by setting $g=\lf\frac{\log \unif{0}{1}}{\log(1-p)}\rf$.

        When the input set does not fit in memory, we discuss our algorithms in the \emph{External Memory} (EM) model~\cite{EM}, the de-facto model for studying I/O-efficient algorithms.
        In EM, a machine is equipped with $M$ words of memory and a disk of an unbounded size that has been formatted into blocks of $B$ words. The values of $M$ and $B$ satisfy $M \geq 2B$. An I/O either reads a block of data from the disk into memory or conversely writes B words from memory into a disk block. The \emph{cost} of an algorithm is defined as the number of I/Os performed (CPU computation is for free), while the \emph{space} of a data structure is the number of blocks occupied.

    \header{\bf Math conventions.} 
        The value of  $\log^*_B n$ is the smallest $t$ such that $$\underbrace{\log_2\log_2\ldots\log_2 n}_t \leq B.$$ When $B=1$, $\log^*_B n$ is the well-known \emph{iterated logarithm} $\log^* n$.
                    
        For a function $\bp$ and a subset $A$ of the domain of $\bp$, $\restr{\bp}{A}$ represents the \emph{restriction} of $\bp$ to $A$, $\bp(A)=\set{\pv|v\in A}$ represents the \emph{image} of set $A$ under function $\bp$.

%% file: sections/introduction/related_work.tex
\subsection{Related Works} \label{sec:related-work}
    \header{\bf Subset sampling.}
        The \emph{subset sampling} problem has been studied for decades, but it has not been fully solved yet. Researchers have made progress in understanding special cases of the problem, such as the case where $\pv=p$ for all $v \in \S$, which is studied in \cite{constant_p_ss}. In this case, they achieve $\bigo{1+\mu_\S}$ expected query time. The general case of the \emph{subset sampling} query is considered in \cite{logn+u_ss}, where researchers design algorithms with $\bigon$ space and $\bigo{\log n + \mu_\S}$ query time. More recently, Bringmann and Panagiotou \cite{1+u_ss} design an algorithm that solves the \emph{subset sampling} problem with $\bigon$ space and $\bigo{1+\mu_\S}$ expected query time. They additionally prove their solution is optimal. 
        It is worth noting that all of the solutions mentioned above consider static dataset that fits in memory. In real-world applications, however, the dataset may change over time or may be too large to be kept in memory.
        
    \header{\bf Independent range sampling.} 
        To the best of our knowledge, no existing work addresses the \emph{range subset sampling} query. The most closely related research to our work is the line of \emph{independent range sampling} (IRS), which is an ``orthogonal''  query of the \emph{range subset sampling} query. IRS aims to sample $k$ elements from the points that fall into the given range. Hu et al.~\cite{irs} study the unweighted case of IRS. They design a dynamic data structure for the scenario where the input set fits in memory, the expected query time is $\bigo{\log n +k}$, the space consumption is $\bigon$ and the \emph{update} time is $\bigo{\log n}$. 
        In \cite{revisited}, the \emph{weighted IRS} (wIRS) is studied. Under the assumption that the key value is defined on $\setht{1,2}{U}$, the query time complexity is improved to $\bigo{Pred(U, w, n)+k}$, where $Pred(U, w, n)$ is the query time of a predecessor search data structure to find the predecessor of $w$ in the binary search tree. It uses $\bigon$ space on an input size $n$ from the universe $\setht{1,2}{U}$ and on a machine with $w$-bit integers~\cite{pred}. In \cite{pods22}, wIRS is solved with $\bigo{\log n +k}$ query time, with no additional assumption. Yet, both \cite{revisited} and \cite{pods22} do not support efficient dynamic maintenance of the data structures. Xie et al.~\cite{sirs} study weighted spatial IRS, they mostly looks at $2d$ rectangle queries and only offer a solution with $\bigo{\sqrt{n} + k}$ query time and $\bigo{\log^2 n}$ update time. 

    \header{\bf Dynamic maintenance of a distribution.}
        There are works that try to design a dynamic structure for sampling. Hagerup et al.~\cite{dynamic_distribution} considers a \emph{distribution} as an abstract data type that represents a probability distribution $f$ on a finite set and supports a \emph{generate} operation, which returns a random value distributed according to $f$ and independent from previous queries. They study the dynamic maintenance of a \emph{distribution}, which supports changes to the probability value through \emph{update} operation. They finally achieve constant expected \emph{generate} time, constant \emph{update} time and $\bigon$ space. Some ideas from this work can be borrowed to design our dynamic data structure, although the problem we study is how to generate a subset instead of a single value. 

%% file: sections/introduction/contributions.tex
\subsection{Contributions} \label{sec:contribution}
    We make three main contributions.
    
    \input{tables/contributions}
          
    \header{\bf A dynamic data structure for SS under the \emph{Real RAM} model.} 
        The key technique is to reduce the original problem of size $n$ to a smaller problem of size $\bigo{\log{n}}$ by a set partition strategy according to the probabilities. We can further reduce the problem size by the above reduction strategy.
        When the problem size becomes sufficiently small, we present a nontrivial table lookup solution that can efficiently answer the \emph{subset sampling} query and support \emph{update} operation on a problem of size $m$ at the cost of increased space, which is $m^{\bigo{m}}$. Then, with the original problem size of $n$ reduced to $\bigo{\log\log n}$, the lookup table can be maintained with $\bigon$ space. This helps achieve our space and time complexity.

    \header{\bf An I/O-efficient algorithm for SS under the \emph{EM} model.} 
        The main challenge in transferring the size reduction technique of the first result to the \emph{EM} model is that this technique relies heavily on \emph{random access}. But \emph{random access} can not be done efficiently in the \emph{EM} model. We tackle this issue by observing that, conditioned on the size of the sample subset is a fixed number, the \emph{subset sampling} reduces to the \emph{set sampling} problem. The \emph{set sampling} problem under the \emph{EM} model is studied and well solved  in~\cite{irs}. It is difficult to transfer the sophisticated table lookup technique to the \emph{EM} model, so we use the size reduction technique repeatedly until the problem becomes small enough to be kept in memory. Then we can end the recursion by using the first result. Finally, our algorithm uses $\bigo{n/B}$ space and achieves $\Bigo{(\log^*_B n)/B+(\mu_\mathcal{S}/B)\log_{M/B} (n/B)}$ amortized expected I/O cost.
        
    \header{\bf A dynamic data structure for RSS under the \emph{Real RAM} model.}
        We utilize the first result and solve the dynamic RSS with $\bigo{\log n+\mu_{\S\cap[a,b]}}$ expected query time, $\bigon$ space, and  $\bigolog$ expected modification time. The key idea is to maintain a binary search tree to deal with the range and then maintain dynamic \emph{subset sampling} structure as a secondary structure at each internal node. This structure shares a similar spirit as existing range sampling structures, like \cite{irs, pods22}. Yet, the key challenge is handling modification operations efficiently with the secondary structure maintained at each internal node. To tackle this issue, we combine the randomized binary search tree \emph{treap} \cite{treap} with our optimal dynamic \emph{subset sampling} structure to achieve the desired time complexity.

        We compare our contributions with related works in Table~\ref{tab:contributions}.

%% file: tables/contributions.tex
\begin{table}[t]\centering
    \caption{Comparison between our results and related works.}
    \label{tab:contributions}
    \scalebox{0.75}{\begin{tabular}{c|c|c|c|c|c} 
        {\bf Problem}& {\bf Computation Model} & {\bf Ref} & {\bf Space} & {\bf Query} & {\bf Modification} \\
        \hline
        \hline 
        SS & \emph{Real RAM} & \cite{1+u_ss} & $\bigon$ & $\bigo{1+\mu_\S}$ & no support  \\
        \hline
        SS & \emph{Real RAM} & this paper & $\bigon$ & $\bigo{1+\mu_\S}$ & $\bigone$  \\
        \hline
        \hline
        IRS & \emph{External Memory}    & \cite{irs} & $\bigo{n/B}$ & $\Bigo{\log^*(n/B)+\log_B n+(k/B)\log_{M/B}(n/B)}$ & no support\\
        \hline
        SS & \emph{External Memory}     & this paper & $\bigo{n/B}$ & $\Bigo{(\log^*_B n)/B+(\mu_\S/B)\log_{M/B}(n/B)}$ & no support  \\
        \hline
        \hline
        wIRS & \emph{Real RAM} & \cite{pods22} & $\bigon$ & $\bigo{\log n+k}$ & no support\\
        \hline
        RSS & \emph{Real RAM} & this paper & $\bigon$ & $\bigo{\log n+\mu_{\S\cap[a,b]}}$ & $\bigolog$  \\
    \end{tabular}}
\end{table}

%% file: sections/dynamic_subset_sampling.tex
\section{A Dynamic Data Structure for SS}\label{sec:dynamic-SS} 
    \input{sections/dynamic_subset_sampling/building_blocks}

    \input{sections/dynamic_subset_sampling/size_reduction}

    \input{sections/dynamic_subset_sampling/table_lookup}

    \input{sections/dynamic_subset_sampling/final_result}

%% file: sections/dynamic_subset_sampling/building_blocks.tex
\subsection{Building Blocks} 
    We first see a straightforward solution.
    
    \input{lemmas/naive}
    
    The query time linear to the problem size in Lemma \ref{naive} is prohibitive in real-world applications. We next show that when $\pv$ for each $v\in\S$ has a tight upper- and lower- bound, i.e., $\bp(\S)\subseteq(2^{-l}, 2^{-l+1}]$ for some fixed $l \geq 1$, we can use geometric random variable to skip some records and reduce query time to $\bigomu$ in expectation without increasing the cost of other operations.
    
    \input{lemmas/geo}

%% file: lemmas/naive.tex
\begin{lemma} [naive solution] \label{naive}
    Given a \emph{subset sampling} problem instance $\Psi=\left<\S, \bp\right>$, we can maintain a data structure $\D$ of size $\bigon$ with $\bigon$ $\que$ time. Besides, $\D$ can be maintained with $\bigone$ $\upd$ time, $\bigone$ amortized expected $\ins$ and $\del$ time.
\end{lemma}

\begin{proof}
    It is easy to achieve the above \emph{update}, \emph{insert} and \emph{delete} time by maintaining a dynamic array $A$ in arbitrary order and a hash table to map from each record $v\in\S$ to its location in array $A$. We further maintain the size $n$ of $\S$. To answer query, for each $j\in\setht{1,2}{n}$, we draw a random number $u_j$ from $\unif{0}{1}$. If $u_j < \bp(A[j])$, we include $A[j]$ in $X$. Since every record is examined once, the query time is thus $\bigon$.
\end{proof}

%% file: lemmas/geo.tex
\begin{lemma}[jump solution for bounded probability]\label{grs} 
    Given a \emph{subset sampling} problem instance $\Psi=\left<\S, \bp\right>$, if $\bp(\S)\subseteq(2^{-l}, 2^{-l+1}]$ for a fixed $l \geq 1$, then we can maintain a data structure $\D$ of size $\bigon$ with $\bigomu$
    expected $\que$ time. Furthermore, $\D$ can be dynamically maintained with $\bigone$ $\upd$ time, $\bigone$ amortized expected $\ins$ and $\del$ time. 
\end{lemma}

\begin{proof} The data structure and modification operations can be handled similar to Lemma \ref{naive}.

    We answer the \emph{subset sampling} query as follows. We first overestimate the probability of each $v\in\S$ as $\bar{p}=2^{-l+1}$. When the probabilities are all equal to $\bar{p}$, we can quickly sample a subset $\bar{X}$ by repeatedly drawing a random number $g$ from $\geo{\bar{p}}$ and then jump to the next $g$ position, and include the record at this position in $\bar{X}$. We keep jumping until we reach the end of the array $A$. Because we may grant $v\in\bar{X}$ with extra probability, we should draw a number $u$ from $\unif{0}{\bar{p}}$ and finally include $v$ in $X$ if and only if $u\leq\pv$. In this way, each record $v\in\S$ is sampled into $X$ with probability $\bar{p}\times\frac{\pv}{\bar{p}}=\pv$. 

    We next analyze the query time.
    Let $g_1, g_2, \ldots$ be the sequence of random number we generate during the above process. Let $t$ be the smallest number such that $\sum_{i=1}^{t+1} g_i\geq n$. On the one hand, $t$ can be interpreted as the number of jumps before reaching the end. On the other hand, $t$ can be interpreted as the number of success trails among $n$ independent trails each with success probability $\bar{p}$. We can take the later interpretation and derive that $t\sim\bin{n}{\bar{p}}$. The expected number of jumps is thus $E[t]=n\bar{p}=\sum_{v\in\S}2\pv=2\mu_\S$. So the expected query time is $\bigo{1+\mu_\S}$.
\end{proof}

%% file: sections/dynamic_subset_sampling/size_reduction.tex
\subsection{Size Reduction} 
    Bringmann and Panagiotou~\cite{1+u_ss} have proved that, using only $\bigon$ space, the optimal query time for the \emph{subset sampling} problem is expected $\bigo{1+\mu_\S}$. Thus the time complexity in Lemma \ref{grs} is essentially optimal. But Lemma \ref{grs} requires bounded probabilities. In general case, the probabilities can be arbitrary values in the interval $[0,1]$. So we divide $\S$ into several baskets such that each basket has bounded probabilities. Since each basket can be maintained efficiently, the problem reduces to sample a subset of the set of baskets. More formally, we have the following definitions:
    
    \input{lemmas/reduce}

%% file: lemmas/reduce.tex
We define $\S_l=\{v|v\in\S,2^{-l}<\pv\leq2^{-l+1}\}(l\in\integer^+)$. Readers can interpret $\S_{l}$ as the basket which stores records whose sampling probabilities fall in the range $(2^{-l},2^{-l+1}]$. 
 
Furthermore, we define $\S_{l^+}=\cup_{i\geq l}\S_i$. Readers can interpret $\S_{l^+}$ as the basket which stores records whose sampling probabilities are less than or equal to $2^{-l+1}$.

Let $Q$ be a subset of $\S$. We define a function $\bq_\bp:\P(\S)\to[0,1]$, where $\P(\S)$ is the power set of $\S$ and $\bq_\bp(Q)=\Pr[$at least one element in $Q$ is sampled$]=1-\prod_{v\in Q}(1-\pv)$. According to the union bound, we have $\bq_\bp(Q)\leq\min{1}{\sum_{v\in Q}\pv}= \min{1}{\mu_Q}$.
        
We define $L=2\lc\log|\S|\rc+2$, which is the number of baskets we will maintain in the Lemma \ref{sr}. We define $\widetilde{\S}=\{1,2,...,L\}$, which is the index set for the baskets. We then define 
$$\widetilde{\bp}(i)=
    \left\{
    \begin{aligned}
    &\bq_\bp(\S_i),1\leq i\leq L-1 \\
    &\bq_\bp(\S_{i^+}) ,i = L
    \end{aligned}\right. 
,$$
which assigns the sampling probability for each basket.

Then, we have the following lemma:

\begin{lemma}[size reduction] \label{sr}
    Given a \emph{subset sampling} problem instance $\Psi=\left<\S, \bp\right>$, 
    assume that using method $\mathcal{M}$, for the \emph{subset sampling} problem instance $\widetilde\Psi=\left<\smS, \sbp\right>$, we can maintain a data structure $\widetilde{\D}$ of size $\bigo{s_0}$ with $\bigo{t_0}$ $\que$ time and $\bigone$ amortized expected $\upd$ time. Then for the \emph{subset sampling} problem instance $\Psi$, we can maintain a data structure $\D$ of size $\bigo{s_0+|\S|}$ with $\bigo{t_0+1+\mu_\S}$ expected $\que$ time and $\bigone$ amortized expected $\upd$, $\ins$ and $\del$ time.
\end{lemma}

\begin{proof}
    The data structure, query operation and modification operations are as follows.

    \header{\bf Structure.}
        Because $\Pr$[at least one record $v\in\S_{\twologtwo^+}$ is sampled$]\leq\frac{1}{2^{-L+1}}\times \norm{\S_{L^+}}\leq\frac{1}{\norm{\S}^2}\times \norm{\S}=\frac{1}{\norm{\S}}$ according to the union bound, we can maintain a data structure $\D_{L^+}$ for $\left<\S_{\twologtwo^+}, \frac{\restr{\bp}{\S_{\twologtwo^+}}}{\widetilde\bp(\twologtwo)}\right>$ using Lemma \ref{naive} (naive solution) without worrying about degrading the overall query time. 
        Here, $\restr{\bp}{\S_{\twologtwo^+}}$ indicates the \emph{restriction} of function $\bp$ to $\S_{\twologtwo^+}$. 
        It is defined for the consistency of the definition of a \emph{subset sampling} problem instance. 
        We maintain data structure $\D_l$ for $\left<\S_l, \frac{\restr{\bp}{\S_l}}{\widetilde\bp(l)}\right>$ using Lemma \ref{grs} (jump solution for bounded probability) for $1\leq l\leq\twologone$. Finally we maintain a data structure $\widetilde{D}$ for $\widetilde\Psi$ using $\mathcal{M}$.
        Note that we store $\bp$ and $\widetilde\bp(l)$ separately instead of storing $\frac{\bp}{\widetilde\bp(l)}$ explicitly for $1\leq l\leq\twologtwo$.
        We also record the size of $\S$ when we first initialize the data structure in $OldSize$ and record the current size of $\S$ in $Size$. 
        The total space used is $\bigo{s_0}+\bigo{\norm{\S}}=\bigo{s_0+\norm{\S}}$.
    
    \header{\bf Query.}
        For \emph{subset sampling} query, we first draw an index set $\sX$ from $\widetilde{\S}$ using $\widetilde{\D}$. Then we draw $X_l$ from basket $S_l$ using $\D_l$ for each $l\in\sX\setminus\{\twologtwo\}$ and draw $X_{l}$ from basket $\S_{\twologtwo^+}$ using $\D_{l^+}$ for each $l\in\sX\cap\{\twologtwo\}$. Finally we return $X=\cup_{l\in\sX}X_l$. 
        The expected time complexity is:
        \begin{align*}
        &\bigo{t_0}+\sbpo{\twologtwo} \Bigo{\left|\S_{\twologtwo^+}\right|}+\sum_{l=1}^{\twologone}\sbpo{l}\Bigo{1+\frac{\mul}{\sbpo{l}}} \\
        &=\quad \bigo{t_0}+\frac{|\S|}{|\S|^2}\mathcal{O}(|\S|)+\sum_{l=1}^{\twologone}\sbpo{l}+\sum_{l=1}^{\twologone}\bigo{\mul} \\
        &=\quad \bigo{t_0}+\bigone+\mathcal{O}(\mu_\S)+\mathcal{O}(\mu_\S) \\
        &=\quad \bigo{t_0+1+\mu_\S}.
        \end{align*}
        In this way, each record $v\in\S_l$ for $1\leq l\leq\twologone$ is sampled into $X$ with probability $\sbpo{l}\times\frac{\pv}{\sbpo{l}}=\pv$. Each record $v\in\S_{\twologtwo^+}$ is sampled into $X$ with probability $\sbpo{\twologtwo}\times\frac{\pv}{\sbpo{\twologtwo}}=\pv$.
    
    \header{\bf Modification.}
        For $\updtwo{v}{p}$ operation, we first locate the $\S_l$ which contains $v$ and the $\S_{l^\prime}$ which would contain $v$ after update. This can be done in $\bigo{1}$ time by calculating 
        $$l=\min{-\lf\log \pv\rf}{\twologtwo}, l^{\prime}=\min{-\lf\log p\rf}{\twologtwo}.$$
        If $l\neq l^\prime$, then we delete $v$ from $\S_l$, insert $v$ into in $\S_{l^\prime}$, and update $\sbpo{l}$ and $\sbpo{l^\prime}$ accordingly. Otherwise, we simply do an update operation in $\S_l=\S_{l^\prime}$ and update $\sbpo{l}$ accordingly. In both cases, the time complexity is amortized expected $\bigo{1}$.
        
        For $\instwo{v}{p}$ operation, we first increment $Size$ by one and augment $\bp$ with $\pv=p$. 
        If $Size<2\times OldSize$, then we locate the $\S_l$ which would contain $v$, insert $v$ into $\S_l$ and update $\sbpo{l}$ accordingly. In this case, the time complexity is amortized expected $\bigo{1}$. If $Size=2\times OldSize$, we reconstruct the whole data structure. As an overflow reconstruction happens after at least $\bigo{|\S|}$ insert operations, and the reconstruction time is $\bigo{|\S|}$, thus in this case the time complexity is also amortized expected $\bigo{1}$.

        For $\delone{v}$ operation, we first decrement $Size$ by one and remove $\pv$.
        If $Size>\frac{OldSize}{2}$, then we locate the $\S_l$ which contains $v$, delete $v$ from $\S_l$ and update $\sbpo{l}$ accordingly. In this case the time complexity is amortized expected $\bigo{1}$. If $Size=\frac{OldSize}{2}$, we reconstruct the whole data structure. As an underflow reconstruction happens after at least $\bigo{|\S|}$ delete operations, and the reconstruction time is $\bigo{|\S|}$, thus in this case the time complexity is also amortized expected $\bigo{1}$.   
\end{proof}

%% file: sections/dynamic_subset_sampling/table_lookup.tex
 \subsection{Table Lookup}
    Observe that, after applying Lemma \ref{sr} twice, the problem size becomes $\bigo{\log\log n}$. This small problem size gives us the opportunity to solve the problem directly by table lookup without further recursion. In particular, given a \emph{subset sampling} problem on set $\S^o$ of size $m$, we design a lookup table solution that achieves $\bigo{1+\mu_{\S^o}}$ expected $\que$ time and $\bigone$ update time by using $m^{\bigo{m}}$ space. Notice that, with the table lookup solution, we only support updates but no insertions/deletions. Yet, according to Lemma \ref{sr}, it suffices for the data structure $\widetilde{\D}$ to support updates only. 
    
    \input{lemmas/lookup}

%% file: lemmas/lookup.tex
\begin{lemma}[table lookup]\label{lookup}
    Given a \emph{subset sampling} problem instance $\Psi^o= \left< \S^o, \bp^o \right>$, where $\S^o=\setht{1,2}{m}$ and the \emph{codomain} of function $\bp^o$ is $\setht{\frac{0}{m^2}, \frac{1}{m^2}}{\frac{m^2-1}{m^2}, \frac{m^2}{m^2}}$,  we can maintain a data structure $\D^o$ of size $m^{\bigo{m}}$ with $\bigo{1+\mu_{\S^o}}$ expected $\que$ time. Furthermore, $\D^o$ can be maintained with $\bigone$ $\upd$ time.
\end{lemma}

Note that we assume discretized probability here for the ease of tabulation. We will show how to discretize in Theorem \ref{sampling structure}. Let us first see an example before delving into the proof.
\begin{example}
     Suppose $\S^o=\{1,2,3,4\}$, let us discuss what we should pre-compute for a specific status of $\bp^o$, say $\bp^o(1)=\frac{3}{16}$, $\bp^o(2)=\frac{7}{16}$, $\bp^o(3)=\frac{4}{16}$, $\bp^o(4)=\frac{5}{16}$. Let $p_{i,j}$ be the entry in the $(i+1)$-th row and $(j+1)$-th column of Table \ref{tab:conceptual}. Then $p_{i,j}$ represents the probability that the last sampled record is $i-1$ and the next successfully sampled record is  $j$. Note that record $0$ and $m+1$ are not real. ``The last sampled record is $0$'' stand for the start of the jump process. ``The next successfully sampled record is $m+1$'' stands for the fact that $i,i+1,\dots,m$ are all not sampled. Conceptually, we can draw a random number $u$ from $\unif{0}{1}$, and find the unique $j\in\setht{i,i+1}{m,m+1}$ with $\sum_{k=i}^{j-1}p_{i,k}<u\leq\sum_{k=i}^{j}p_{i,k}$ and decide to jump to $j$ on the fly, which costs $\bigo{m}$ time and is costly. To avoid such expensive online cost, we encode such information with a pre-computed array for $j$ such that it considers all possible statuses of generated random numbers. Observe that after multiplying $p_{i,i},p_{i,i+1},\dots,p_{i,m+1}$ by their lowest common denominator, which is $(m^2)^{m-i+1}$, it suffice to draw a random integer $du$ from $[1,(m^2)^{m-i+1}]$, for deciding where to jump to. Then for fixed $\bp^o$ and $i$, we can easily pre-compute an array of size $(m^2)^{m-i+1}$, where the $k$-th entry of the array stores the unique $j$ we should jump to if the generated random integer $du=k$. We show the array pre-computed for the above $\bp^o$ and $i=2$ in Table \ref{tab:array_2}, the array pre-computed for the above $\bp^o$ and $i=3$ in Table \ref{tab:array_3}.
\end{example}

Next, we are ready to give the detailed proof.
\begin{proof}
    The data structure, query operation, and update operation are as follows.
    
    \input{tables/conceptual}

    \input{tables/array_2}

    \input{tables/array_3}   
    
    \header{\bf Structure.}
        We use a numeral system with radix $m^2+1$ to encode $\textbf{p}^o$, i.e. we represent $\textbf{p}^o$ as: 
        $$\lambda=\left(\left(m^2p^o(m)\right)\ \left(m^2p^o(m-1)\right)\dots\left(m^2p^o(2)\right)\left(m^2p^o(1)\right)\right)_{m^2+1} .$$ 
        Then there are $(m^2+1)^m$ possible statuses of $\textbf{p}^o$, i.e., each number in the numeral system with radix $m^2+1$ indicates a status. For a fixed $\textbf{p}^o$, we calculate 
        $$\beta[\lambda][i]=m^2p^o(i), \overline{\beta}[\lambda][i]=m^2(1-p^o(i)),$$ 
        for $i\in[1,m]$. We set $\beta[\lambda][m+1] = 1$. For each $1\leq i\leq j\leq m$, we pre-compute the integers $Mass[\lambda][i][j]$ that are proportional to the probability that the last sampled record is $i-1$ and the next successfully sampled record is  $j$. Note that when $i=1$, ``the last sampled record is $0$''  stands for the start of the sampling process. We also pre-compute $Mass[\lambda][i][m+1]$, which is proportional to the probability that the records $i,i+1\ldots,m$ are all not sampled. This is done by calculating 
        $$Mass[\lambda][i][j]=\left(\prod_{k=i}^{j-1}\overline{\beta}[\lambda][k]\right)\beta[\lambda][j](m^2)^{\max{0}{m-j}}$$ 
        for $j\in[i+1,m+1]$ and calculating 
        $$Mass[\lambda][i][i]=\beta[\lambda][i](m^2)^{\max{0}{m-i}}.$$ 
        Next, we maintain an array of size $(m^2)^{m-i+1}$ at $Tb[\lambda][i]$. For each $1\leq r\leq{(m^2)^{m-i+1}}$, we fill in $Tb[\lambda][i][r]$ with the unique $j\in\setht{i,i+1}{m,m+1}$ with 
        $$\sum_{k=i}^{j-1}Mass[\lambda][k]<r\leq\sum_{k=i}^{j}Mass[\lambda][k]$$
        in $\mathcal{O}(m)$ time. The total space is thus 
        $$(m^2+1)^m\times m \times m^{\bigo{2m}}=m^{\bigo{m}}.$$ 
        Since each entry can be computed in $\bigo{m}$ time, the initialization time is also $m^{\bigo{m}}$.
    
    \header{\bf Query.}
        For the \emph{subset sampling} query, we start with $i=1$ and draw a random integer $du$ from $[1,(m^2)^{m-i+1}]$. Then, we check the table and access the array stored at table entry $Tb[\lambda][i]$ to determine the next successfully sampled record, suppose $Tb[\lambda][i][du]=j$, then we set $i=j$. This process continues until $m+1$ is sampled, and the expected time complexity is $\Bigo{1+\sum_{v\in\S^o}\pv}=\bigo{1+\mu_{\S^o}}$. According to the construction of the lookup table, each record $v\in\S^o$ is sampled into with correct probability $\bp^o(v)$.
    
    \header{\bf Update.}
         For $\updtwo{v}{p}$ operation, it is equivalent to updating the $v$-th digit of the $(m^2+1)$-based number $\lambda$, which is a generalized \emph{bit operation}. Specifically, we update $\lambda$ to $$\lfloor\frac{\lambda}{(m^2+1)^v}\rfloor(m^2+1)^v+(m^2p)(m^2+1)^{v-1}+\lambda\%(m^2+1)^{v-1},$$ which can be done in $\bigone$ time.
\end{proof}

%% file: tables/conceptual.tex
\begin{table}[t]\centering
    \caption{Conceptual Jump Table}
    \label{tab:conceptual}
    \scalebox{0.8}{\begin{tabular}
    {|c|c|c|c|c|c|}
        \hline \diagbox{i}{Prob.}{j} 
            & 1            & 2            & 3             & 4                 & 5
        \\\hline
        1   & $\pa$        & $\bpa\ti\pb$    & $\bpa\ti\bpb\ti\pc$ & $\bpa\ti\bpb\ti\bpc\ti\pd$ & $\bpa\ti\bpb\ti\bpc\ti\bpd$
        \\\hline
        2   & \diagbox{}{} & $\pb$        & $\bpb\ti\pc$     & $\bpb\ti\bpc\ti\pd$     & $\bpb\ti\bpc\ti\bpd$
        \\\hline
        3   & \diagbox{}{} & \diagbox{}{} & $\pc$         & $\bpc\ti\pd$         & $\bpc\ti\bpd$ 
        \\\hline
        4   & \diagbox{}{} & \diagbox{}{} & \diagbox{}{} & $\pd$              & $\bpd$
        \\\hline
    \end{tabular}}
\end{table}

%% file: tables/array_2.tex
\begin{table}[t]\centering
    \caption{The array pre-computed for the above $\bp^o$ and $i=2$}
    \label{tab:array_2}
    \scalebox{0.85}{\begin{tabular}{ccc}
      \toprule
        Index & $[1, 7\times16^2=1792 ]$ & $[1793,1792 +9\times4\times16=2368]$ \\
      \midrule
        Value & 2 & 3 \\
         \midrule
            Index & $[2369, 2368+9\times12\times5=2908] $ & $[2909, 16^3=4096]$\\
      \midrule
        Value &  4 & 5\\
      \bottomrule
    \end{tabular}}
\end{table}

%% file: tables/array_3.tex
\begin{table}[!t]\centering
    \caption{The array pre-computed for the above $\bp^o$ and $i=3$}
    \label{tab:array_3}
    \scalebox{0.85}{\begin{tabular}{cccc}
        \toprule
        Index & $[1, 4\times16=64]$ & $[65,64 +12\times5=124]$ & $[125, 16^2=256] $
        \\\midrule
        Value  & 3 & 4 & 5
        \\\bottomrule
    \end{tabular}}
\end{table}

%% file: sections/dynamic_subset_sampling/final_result.tex
\subsection{Final Result} 
    After having established the above lemmas, we next show our final solution for the \emph{subset sampling} problem.
    \begin{theorem}\label{sampling structure}
        Given a \emph{subset sampling} problem instance $\Psi=\left<\S, \bp\right>$, we can maintain a data structure $\D$ of size $\mathcal{O}(|\S|)$ with $\bigo{1+\mu_\S}$ $\que$ time. Furthermore, $\D$ can be dynamically maintained with $\bigone$ amortized expected $\upd$, $\ins$ and $\del$ time.
    \end{theorem}
    
    \begin{proof}
        We first apply Lemma \ref{sr} twice to get a \emph{subset sampling} problem instance $\hat{\Psi}=\left<\hat{\S}, \hat{\bp}\right>$, where $m=|\hat{\S}|$ would be $\mathcal{O}(\log\log|\S|)$. We next show how to solve $\hat{\Psi}$.
        
        \header{\bf Structure.}
            We set $\ppv=\frac{\lfloor m^2\phv\rfloor}{m^2}, \pppv=\phv-\ppv$. Define $\Psi^\prime=\left<\hat{\S},\frac{\bp^\prime}{\textbf{q}_{\bp^\prime}(\hat{\S})}\right>, \Psi^{\prime\prime}=\left<\hat{\S},\frac{\bp^{\prime\prime}}{\textbf{q}_{\bp^{\prime\prime}}(\hat{\S})}\right>$. Note that we store $\bp^\prime$ and $\textbf{q}_{\bp^\prime}(\hat{\S})$ (resp. $\bp^{\prime\prime}$ and $\textbf{q}_{\bp^{\prime\prime}}(\hat{\S})$) separately instead of storing $\frac{\bp^\prime}{\bq_{\bp^\prime}(\hat{\S})}$ (resp. $\frac{\bp^{\prime\prime}}{\textbf{q}_{\bp^{\prime\prime}}(\hat{\S})}$) explicitly. We maintain a data structure $\D^\prime$ for $\Psi^\prime$ using Lemma \ref{lookup} (table lookup solution) and maintain a data structure $\D^{\prime\prime}$ for $\Psi^{\prime\prime}$ using Lemma \ref{naive} (naive solution). 
            When $m=\bigo{\log\log|\S|}$, $$m^{\bigo{m}}=\bigo{2^{\log\log |\S|(\log\log\log |\S|)}}=\bigo{|\S|}.$$ Thus the space and initialization time for maintaining $\Psi^\prime$ and $\Psi^{\prime\prime}$ add together to $\bigo{|\S|}$. 
        
        \header{\bf Query.}
            We first draw a random number $u^\prime$ from $\unif{0}{1}$. If $u^\prime<\textbf{q}_{\bp^\prime}(\hat{\S})$, we sample $X^\prime$ from set $\hat{\S}$ using $\D^\prime$, otherwise we set $X^\prime=\emptyset$. Again, we draw a number $u^{\prime\prime}$ from $\unif{0}{1}$. If $u^{\prime\prime}<\textbf{q}_{\bp^{\prime\prime}}(\hat{\S})$, we sample $X^{\prime\prime}$ from set $\hat{\S}$ using $\D^{\prime\prime}$, otherwise we set $X^{\prime\prime}=\emptyset$. We then return $X^{\prime}\cup X^{\prime\prime}$ as the query result. 
        
            Since $\bp^\prime<\hat{\bp}$, the query time of $\Psi^\prime$ is $\bigo{1+\mu_{\hat{\S}}}=\bigo{1+\mu_\S}$. The query time of $\Psi^{\prime\prime}$ is $\bigo{m}$ but with a probability bounded by $\bigo{1/m}$. The total expected time complexity is:
            \begin{align*}  
                &\bigone+\qa\Bigo{1+\frac{\mu_\S}{\qa}}+\qb\bigo{m}\\
                =\quad& \bigo{1}+\bigo{1+\mu_\S}+m\frac{1}{m^2}\bigo{m}\\
                =\quad& \bigo{1}+\bigo{1+\mu_\S}+\bigone \\
                =\quad& \mathcal{O}(1+\mu_\S).
            \end{align*}
            In this way, each $v\in\hat{\S}$ is sampled with probability $$\qa\frac{\bp^\prime(v)}{\qa}+\qb\frac{\bp^{\prime\prime}(v)}{\qb}=\bp^\prime(v)+\bp^{\prime\prime}(v)=\hat{\bp}(v).$$
        
        \header{\bf Modification.}
            For $\updtwo{v}{p}$ operation, we simply do an \emph{update} operation in $\Psi^\prime$ and then another \emph{update} operation in $\Psi^{\prime\prime}$. The $\qa$ and $\qb$ we maintained can also be updated in $\bigone$ time. The overall \emph{update} cost of $\hat{\Psi}$ is thus $\bigone$.
        
        Having established the above results  for $\hat{\Psi}$, we can apply Lemma \ref{sr} twice to get the results stated in this theorem.
    \end{proof}

%% file: sections/em.tex
\section{An I/O-efficient Algorithm for SS}\label{sec:em}
    \input{sections/em/building_blocks}

    \input{sections/em/size_reduction}

    \input{sections/em/final_result}

%% file: sections/em/building_blocks.tex
\subsection{Building Blocks in \emph{EM}}
    We first discuss the \emph{EM} version of Lemma \ref{naive} and Lemma \ref{grs}.
    
    \input{lemmas/em_naive}

    Unfortunately, Lemma \ref{grs} involves \emph{random access}, it is nontrivial to transfer Lemma \ref{grs} into \emph{EM}. Before investigating the \emph{EM} version of Lemma \ref{grs}, let us first introduce the \emph{set sampling} problem proposed by Hu et al. in~\cite{irs}.

    \begin{problem}[Set Sampling~\cite{irs}]
        Pre-process a set $\S$ of $n$ elements such that all queries of the following form can be answered efficiently: given an integer $s\geq1$, randomly sample $s$ element \emph{with replacement} or \emph{without replacement} from $\S$.
    \end{problem}

    Then, Hu et al.~\cite{irs} prove the optimal I/O cost of $\Bigo{\min{s}{(s/B)\log_{M/B} (n/B)}}$ for the \emph{set sampling} problem. They also propose an optimal set sampling structure with $\Bigo{(s/B)\log_{M/B} (n/B)}$ amortize I/O cost and $\bigo{n/B}$ space. Note that they consider only the case when $(1/B)\log_{M/B} (n/B)\leq 1$, otherwise it suffices to run the RAM structure in EM directly. Using this structure, each sampled element is charged only $\Bigo{(1/B)\log_{M/B} (n/B)}$ I/Os. 

    Now we are ready to discuss the \emph{EM} version of Lemma \ref{grs}.
    
    \input{lemmas/em_geo}

%% file: lemmas/em_naive.tex
\begin{lemma} [naive solution in EM] \label{em_naive} 
    Given a \emph{subset sampling} problem instance $\Psi=\left<\S, \bp\right>$, there is a data structure $\D$ of $\bigo{n/B}$ space that answers a query in $\bigo{n/B}$ amortized expected I/Os.
\end{lemma}

\begin{proof}
    Since the naive solution only involves array scanning, the space and query time under the \emph{EM} model are that of the \emph{Real RAM} model divided by $B$.
\end{proof}

%% file: lemmas/em_geo.tex
\begin{lemma}[jump solution for bounded probability in \emph{EM}]\label{em_grs} 
    Given a \emph{subset sampling} problem instance $\Psi=\left<\S, \bp\right>$, if $\bp(\S)\subseteq(2^{-l}, 2^{-l+1}]$ for a fixed $l \geq 1$, then there is a data structure $\D$ of $\bigo{n/B}$ space that answers a query in $\Bigo{(\mu_\S/B)\log_{M/B} (n/B)}$ amortized expected I/Os.
\end{lemma}

\begin{proof} 
The data structure and the query method are as follows.

\header{\bf Structure.}
    We use the set sampling structure in~\cite{irs} to store $\S$.

\header{\bf Query.}
    We first overestimate the probability of each $v\in\S$ as $\bar{p}=2^{-l+1}$. Then we keep drawing random number $g_i$ from $\geo{\bar{p}}$ ($i=1,2,\ldots$). Let $t$ be the smallest number such that $\sum_{i=1}^{t+1} g_i\geq n$. We stop drawing random numbers immediately after $g_{t+1}$ is drawn. 
    Notice that this process can be done in memory with no extra I/Os. 
    Then we initiate a \emph{without replacement} query with parameter $t$ to the set sampling structure. Let $Z$ denote the random subset returned by the set sampling structure. We define $\bar{\bp}(v)=\bar{p}$ for $v\in\S$. Recap that $t\sim B(n,\bar{p})$. Then we have
    \begin{align*}
    \Pr[Z=T\subseteq\S]&=\Pr[t=\norm{T}]\cdot\Pr[Z=T\mid t=\norm{T}]\\
    &=\tbinom{n}{\norm{T}}\bar{p}^{\norm{T}}(1-\bar{p})^{n-\norm{T}}\cdot\frac{1}{\tbinom{n}{\norm{T}}}\\
    &=\bar{p}^{\norm{T}}(1-\bar{p})^{n-\norm{T}}\\
    &=\left( \prod_{v\in T}\bar{\bp}(v) \right)\left( \prod_{v\in {\S\setminus T}}(1-\bar{\bp}(v)) \right).
    \end{align*}
    Thus $Z$ can be regarded as a query result from \emph{subset sampling} problem instance $\left<\S, \bar{\bp}\right>$.

    Since the records in $Z$ are sampled with overestimated probability, we need to scan across $Z$ and draw a random number $u$ from $\unif{0}{\bar{p}}$ for each $v\in Z$. If $u<\pv$, we keep $u$, otherwise we discard $u$. After this checking process, we finally get a valid result for \emph{subset sampling} query on $\Psi$.
    
    Recap that the expectation of $t$ is $E[t]=n\times\bar{p}\leq\sum_{v\in\S}2\pv=2\mu_\S$, so the amortized expected query time is:
    $$E[t]\Bigo{(1/B)\log_{M/B} (n/B)}=\Bigo{(\mu_\S/B)\log_{M/B} (n/B)}.$$
\end{proof}

%% file: sections/em/size_reduction.tex
\subsection{Size Reduction in \emph{EM}} 
    It is relatively straightforward to transfer Lemma \ref{sr} to \emph{EM}. We replace Lemma \ref{naive} and Lemma \ref{grs} used in Lemma \ref{sr} with their \emph{EM} versions to get the following lemma.
        
    \input{lemmas/em_reduce}

%% file: lemmas/em_reduce.tex
\begin{lemma}[size reduction in \emph{EM}] \label{em_sr}
    Given a \emph{subset sampling} problem instance $\Psi=\left<\S, \bp\right>$, assume that using method $\mathcal{M}$, for the \emph{subset sampling} problem instance $\widetilde\Psi=\left<\smS, \sbp\right>$, there is a data structure $\widetilde{\D}$ of $\bigo{s_0}$ space that answers a query in $\bigo{t_0}$ I/Os, then for the \emph{subset sampling} problem instance $\Psi$, there is a data structure $\D$ of $\bigo{s_0+|\S|/B}$ space that answers a query in $\bigo{t_0+1/B+(\mu_\S/B)\log_{M/B} (\norm{\S}/B)}$ amortized expected I/Os.
\end{lemma}

\begin{proof}
    The data structure and the query method are as follows.

    \header{\bf Structure.}
        We now maintain a data structure $\D_{L^+}$ for $\left<\S_{\twologtwo^+}, \frac{\restr{\bp}{\S_{\twologtwo^+}}}{\widetilde\bp(\twologtwo)}\right>$ using Lemma \ref{em_naive} (naive solution in \emph{EM}), maintain data structure $\D_l$ for $\left<\S_l, \frac{\restr{\bp}{\S_l}}{\widetilde\bp(l)}\right>$ using Lemma \ref{em_grs} for $1\leq l\leq\twologone$. Finally we maintain a data structure $\widetilde{\D}$ for $\widetilde\Psi$ using $\mathcal{M}$. Since we focus on static structure in \emph{EM} and we don't need to support dynamic operation in this case, $\frac{\bp}{\widetilde\bp(l)}$ ($1\leq l\leq\twologtwo$) can be stored explicitly. The total space used is $\bigo{s_0}+\bigo{\norm{\S}}=\bigo{s_0+\norm{\S}}$.
    
    \header{\bf Query.}
        The steps for answering the query are the same as that of Lemma \ref{sr}. The amortized expected time complexity:
        \begin{align*}
        &\bigo{t_0}+\sbpo{\twologtwo} \Bigo{\left|\S_{\twologtwo^+}\right|/B}+\sum_{l=1}^{\twologone}\sbpo{l}\Bigo{\left(\frac{\mu_{\S_l}}{\sbpo{l}B}\right)\log_{M/B} (n/B)} \\
        &=\quad \bigo{t_0}+\frac{|\S|}{|\S|^2}\bigo{|\S|/B}+\sum_{l=1}^{\twologone}\Bigo{(\mu_{\S_l}/B)\log_{M/B} (n/B)} \\
        &=\quad \bigo{t_0}+\bigo{1/B}+\Bigo{(\mu_\S/B)\log_{M/B} (n/B)}\\
        &=\quad \Bigo{t_0+1/B+(\mu_\S/B)\log_{M/B} (n/B)}.
        \end{align*}
\end{proof}

%% file: sections/em/final_result.tex
\subsection{Final Result in \emph{EM}}
    The table lookup solution in Lemma \ref{lookup} is sophisticated and relies heavily on \emph{random access}, so it is difficult to transfer it to \emph{EM} efficiently. Thus we use the size reduction technique in \emph{EM} repeatedly until the problem become small enough to be put in memory.
    \begin{theorem}\label{sampling structure in EM}
        Given a \emph{subset sampling} problem instance $\Psi=\left<\S, \bp\right>$, there is a data structure $\D$ of $\bigo{n/B}$ space that answers a query in $\Bigo{\frac{\log^*_B n}{B}+(\mu_\S/B)\log_{M/B} (n/B)}$ amortized expected I/Os. 
    \end{theorem}
    
    \begin{proof}
        We repeatedly use Lemma \ref{em_sr} until the problem size becomes less than $B$ so that we can use Theorem \ref{sampling structure} in memory to end the recursion. Thus there will be $\log^*_B n$ recursions with decreasing problem sizes. We next analyze the space and time complexity. The space usage (the number of blocks occupied) is:
        $$
            \underbrace{\bigo{n/B}+\bigo{\log n/B}+\bigo{\log\log n/B}+\ldots+\bigo{1}}_{\log^*_B n}=\bigo{n/B}.
        $$
        The amortized expected I/O cost is:
        \begin{align*}
            &\underbrace{\Bigo{\frac{1}{B}+(\frac{\mu_\S}{B})\log_{M/B} (n/B)}+\Bigo{\frac{1}{B}+(\frac{\mu_\S}{B})\log_{M/B} (\log n/B)}+\ldots}_{\log^*_B n}\\
            &=\Bigo{(\log^*_B n)/B+(\mu_\S/B)\log_{M/B} (n/B)}.
        \end{align*}
    \end{proof}

%% file: sections/dynamic_range_subset_sampling.tex
\section{A Dynamic Data Structure for RSS} 
    We will next show how to design a dynamic data structure for the \emph{range subset sampling} problem with the structure proposed in Section \ref{sec:dynamic-SS}. We will first design a data structure with $\bigo{\log n+\mu_{\S\cap[a,b]}}$ amortized expected query time, $\bigolog$ expected modification time, and $\bigonlog$ space in Section \ref{sec:rss_baseline}. Then, we improve it to using only $\bigon$ space in Section \ref{sec:rss_on}.
    
    \input{sections/dynamic_range_subset_sampling/a_baseline_solution}

    \input{sections/dynamic_range_subset_sampling/final_result_with_linear_space}

%% file: sections/dynamic_range_subset_sampling/a_baseline_solution.tex
\subsection{A baseline solution}\label{sec:rss_baseline}        
    \input{lemmas/rssp_baseline}

%% file: lemmas/rssp_baseline.tex
Before giving our solution, we first introduce \emph{treap}~\cite{treap}, an important data structure that will be used in our solution.

\begin{definition}[Treap\cite{treap}]
    Given a set $P$ of $n$ items where each item is associated with a \emph{key} and a \emph{priority}. A \emph{treap} for set $P$ is a rooted binary tree $\T$ with node set $P$, arranged in in-order with respect to the keys and in heap-order with respect to the priorities. In-order means that for any node $x$ in the tree, $y.key \leq x.key$ (resp. $y.key \geq x.key$) for all $y$ in the left (resp. right) subtree of $x$. Heap-order means that for any node $x$ with parent $p$, the relation $x.priority < p.priority$ holds.
\end{definition}

The \emph{treap} data structure has several properties which make it a useful tool for designing nested dynamic data structures.
\begin{theorem}[properties of treap\cite{treap}]
    If $\T$ is a treap for $P$ where the priorities of the items are independent, identically distributed random variables, then the following holds:
    \begin{itemize}
        \item The expected time  to access an item $x\in P$ in tree $\T$ is $\bigolog$.
        \item The expected time  to perform an insertion into $\T$ or a deletion from $\T$ is $\bigolog$.
        \item If the cost of a rotation (including the update cost of secondary structures associated with tree nodes) to a subtree of size $s$ is $\bigo{s}$. Then, the expected time needed to update the treap for an insertion or deletion is $\bigolog.$ 
        \end{itemize}      
\end{theorem}

\noindent For ease of exposition, we introduce some additional definitions and notations. Let $r$ be the root of $\T$.      
For a node $u$ in $\T$, we define its $slab$ recursively as follows: 
    \begin{itemize}
        \item If $u=r$, then $slab(r)=(-\infty,\infty)$.
        \item Else, let $p$ be the parent of $u$. If $u$ is the left (resp. right) child of $p$, then $slab(u)=slab(p) \cap \left(-\infty, key(p)\right)$ (resp. $slab(p) \cap [key(p),\infty)$).
    \end{itemize}

\begin{figure}[t]
    \centering
    \includegraphics[height=37mm]{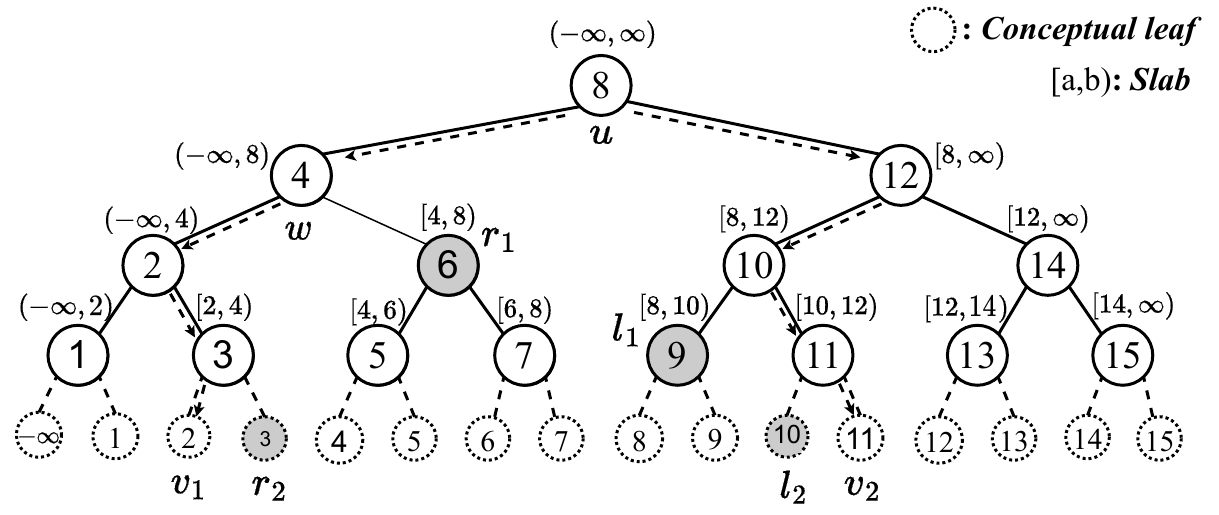}
    \caption{An example treap (priorities omitted)}
    \label{fig:treap}
\end{figure}

\noindent For each node $u\in\T$, we define $Leaves(u)=P\cap slab(u)$. As a well-known fact~\cite{pods22}, for a range $[a,b]$, we can identify a set $\C$ of $\bigo{\log n}$ \emph{canonical nodes} in $\T$ such that: 
    \begin{itemize}
        \item The nodes in $\C$ have disjoint subtrees.
        \item The $Leaves$ in the subtree of the nodes in $\C$ constitute $P\cap[a,b]$.
    \end{itemize}
\noindent See below for an example of \emph{treap}.
\begin{example}\label{treap-example}
    We show an example of \emph{treap} for $P=\setht{1,2}{15}$ in Figure \ref{fig:treap}. Note that the $Leaves$ of a real node are represented with dashed circles. The \emph{slab} of $w$ is $(-\infty,8)$. For node $r_1$, $Leaves(r_1)=\{4,5,6,7\}$. Suppose $[a,b]=[2.5,10.6]$, then the \emph{canonical nodes} of $[a,b]$ is $\C=\{r_1,r_2,l_1,l_2\}$. It follows that $Leaves(r_1) \cup Leaves(r_2)\cup Leaves(l_1)\cup Leaves(l_2)=\{3,4,5,6,7,8,9,10\}$, which is exactly $P\cap[a,b]$.
\end{example}

Now we are ready to describe our algorithm.
        
\header{\bf Structure.}
    On the top level, we build a \emph{treap} $\T$ on $\S$. The \emph{priority} for each newly inserted treap node is drawn from $\unif{0}{1}$ independently. Thus we can conveniently use the properties of \emph{treap}. For each node $u$ in $\T$, 
    we associate $u$ with a dynamic \emph{subset sampling} structure for $Leaves(u)$ using the data structure in Theorem \ref{sampling structure}.
    Since each value $v$ will be stored in the subset sampling structure of $\bigo{h}$ nodes, and $h$ is $\bigolog$ in expectation, the total space of the structure is thus $\bigo{n\log n}$ in expectation.
            
\header{\bf Query.}
    Given input range $[a,b]$ at query time, we answer the query as follows. We first find a set $\C$ of \emph{canonical nodes} for $[a,b]$ in $\bigolog$ time. Suppose $\C=\setht{u_1, u_2}{u_c}$ for some $c=\bigolog$. 
    We sample a subset $X_i$ from the dynamic \emph{subset sampling} structure of $u_i$ for $i\in\setht{1}{c}$.
    Finally we return $\cup_{i=1}^{c}X_c$. The overall expected query time is: $$\bigolog+\sum_{i=1}^{c}\bigo{1+\mu_{Leaves(u_i)}}=\bigo{\log n+\mu_{\S\cap[a, b]}}.$$
            
\header{\bf Modification.}   
    For $\instwo{v}{p}$ operation, we insert $v$ into $\T$ in a two-phase manner. We first insert $v$ into $\T$ according to its key value and update the secondary structures of the nodes on the inserting path. Since the cost of updating the secondary structure of each node is $\bigone$, the time complexity of the first phase is expected $\bigolog$. In the second phase, we may need to rotate up $v$ if the priority generated for it is high. Since our dynamic \emph{subset sampling} structure supports constant time modification, the cost of a rotation of a subtree of size $s$ is $\bigo{s}$. Thus according to the properties of treap, we know that the expected time to perform an insertion to $\T$ and updating its secondary structures is $\bigolog$. 
    
    For $\delone{v}$ operation, we first rotate $v$ down to the child with greater priority repeatedly until $v$ becomes a leaf and update the secondary structures of the nodes on the path. Finally we remove this leaf. The other steps and analysis for $\delone{v}$ are analogous to $\instwo{v}{p}$ operation.
    
    From this point forward, we will implement $\updtwo{v}{p}$ operation by first performing $\delone{v}$, followed by $\instwo{v}{p}$.

Now we can derive the following lemma.
\begin{lemma} \label{rssp_baseline}
    Given a \emph{range subset sampling} problem instance $\psi=\left<\S, \bp\right>$, we can maintain a data structure $\D$ of size $\bigo{|\S|\log |\S|}$ with $\bigo{\log |\S|+\mu_{\S\cap[a,b]}}$ expected $\que$ time if the input range is $[a,b]$. Furthermore, $\D$ can be dynamically maintained with $\bigo{\log |\S|}$ amortized expected $\upd$, $\ins$ and $\del$ time.
\end{lemma}

%% file: sections/dynamic_range_subset_sampling/final_result_with_linear_space.tex
\subsection{Final result with linear space}\label{sec:rss_on} 
    We next show how to reduce the space complexity to $\bigon$. 
        
    \input{lemmas/rssp_final}

%% file: lemmas/rssp_final.tex
\header{\bf Structure.}
    Let $s$ be an integer between $(\log n - 1)$ and $(\log n + 1)$. We divide $\real$ into a set of $g=\Theta(\frac{n}{\log n})$ disjoint intervals $\I_1, ..., \I_g$ such that each $\I_i(1\leq i\leq g)$ covers between $\frac{s}{2}$ and $s$ records of $\S$. Then we define $\C_i=\I_i\cap\S$ as the $i$-th $chunk$. 
        
    We first create a structure $\T_{chunk}$ on the chunk level using Lemma \ref{rssp_baseline}. More precisely, the key set is $S_{chunk}=\{1,2,...,g\}$ and the probability function is $\bp_{chunk}(i)=\textbf{q}_\bp(\C_i)=1-\prod_{v\in\C_i}(1-\pv)$. The space of $\T_{chunk}$ is $\bigo{g\log g}=\bigo{\frac{n}{\log n}\log \frac{n}{\log n}}=\bigon$. We also build a dynamic \emph{subset sampling} structure for each $\C_i(1\leq i\leq g)$ using Theorem \ref{sampling structure}. Note that the sampling probability of each $v\in\C_i$ should be scaled to $\frac{\bp(v)}{\bp_{chunk}(i)}$. The \emph{subset sampling} structures of all chunks use $\bigo{\sum_{i=1}^g \norm{\C_i}}=\bigon$ space in total. Thus the overall space consumption of our structure is $\bigon$.
        
\header{\bf Query.}
    Given input range $[a,b]$ at query time, we answer the \emph{range subset sampling} query as follows. We first locate the intervals $\I_l$ and $\I_r$ that contain $a$ and $b$, respectively. If $l=r$, we can answer the query by brute force, i.e. scan $\C_l=\C_r$ and decide if each record is sampled or not, this can be done in $\bigolog$ time.

    If $l\neq r$, we can spend $\bigolog$ time to scan the marginal chunks $\C_l$ and $\C_r$ to decide if each record is in the range and is sampled or not. For the chunks sandwiched by $\C_l$ and $\C_r$, we first query $\T_{chunk}$ with range $[l+1,r-1]$ to get an index subset $I$. Then, for each $i\in I$, we draw a subset $X_i$ using the \emph{subset sampling} structure of $\C_i$. Finally, we return $\cup_{i\in I}X_i$. The expected time is:
    \begin{align*}
        &\bigolog+\sum_{i=l+1}^{r-1}\bp_{chunk}(i)\mathcal{O}(1+\frac{\mu_{\C_i}}{\bp_{chunk}(i)})\\
        =\quad&\bigolog+\sum_{i=l+1}^{r-1}\bigo{\bp_{chunk}(i)}+\sum_{i=l+1}^{r-1}\bigo{\mu_{\C_i}}\\
        =\quad&\bigolog+\bigo{\sum_{v\in\S\cap[a,b]}\pv}+\bigo{\mu_{\S\cap[a,b]}}\\
        =\quad&\bigolog+\bigo{\mu_{\S\cap[a,b]}}+\bigo{\mu_{\S\cap[a,b]}}\\
        =\quad&\bigo{\log n+\mu_{\S\cap[a,b]}}.
    \end{align*}
    Thus, the query time complexity is $\bigo{\log n + \mu_{\S\cap[a,b]}}$.
        
\header{\bf Modification.} 
    $\T_{chunk}$ is updated whenever the number of points of a chunk changes. This can be done in $\bigo{\log \frac{n}{\log n}}=\bigolog$ time per $\ins\ (\del)$ operation of a point in $\S$. When a chunk overflows (resp. underflows), it can be repaired in $\bigo{s}$ time by chunk split (resp. merge). Using standard amortization analysis, each modification operation only bears $\bigone$ time amortized. Finally, when the size of $\S$ doubles or halves, we rebuild the whole data structure to make sure $s$ is between $(\log n-1)$ and $(\log n  +1)$, and reset $s$ according to the new $n$. Thus the overall modification cost is amortized expected $\bigolog$.

Now we can establish the following theorem.
\begin{theorem}\label{rssp_final}
    Given a \emph{range subset sampling} problem instance $\psi=\left<\S, \bp\right>$, we can maintain a data structure $\D$ of size $\bigon$ with $\bigo{\log n+\mu_{\S\cap[a,b]}}$ expected $\que$ time if the input range is $[a,b]$. Furthermore, $\D$ can be dynamically maintained with $\bigolog$ amortized expected $\upd$, $\ins$ and $\del$ time.
\end{theorem}

%% file: sections/conclusion.tex
\section{Conclusion}
    The \emph{subset sampling} problem is widely studied and has many applications in databases, e.g., ~\cite{constant_p_ss, logn+u_ss, 1+u_ss}. For this problem, we first present a dynamic data structure in the \emph{Real RAM} model, which supports modification operations in $\bigone$ time while achieving optimal expected query time $\bigo{1+\mu_\S}$ and space complexity $\bigon$. We next describe an I/O-efficient data structure in the \emph{EM} model, which uses $\bigo{n/B}$ space and answers a query with $\Bigo{(\log^*_B n)/B+(\mu_\S/B)\log_{M/B} (n/B)}$ amortized expected I/Os. Finally, we extend the \emph{subset sampling} problem to the \emph{range subset sampling} problem and designed a nontrivial solution for it in the \emph{Real RAM} model. The query time is amortized expected $\bigo{\log n + \mu_\S}$, the modification time is amortized expected $\bigolog$, the space consumption is $\bigon$.
    Future research directions may include considering higher dimensional case of SS, external memory case of RSS, etc.